\documentclass[10pt,letterpaper,twocolumn]{article} 

\usepackage{ol2}
\usepackage[draft,implicit=false]{hyperref}
\usepackage{amsmath}

\usepackage{amssymb}

\newcommand{\GH}{Goos-H\"anchen }
\newcommand{\IF}{Imbert-Fedorov }
\newcommand{\vett}[1]{\mathbf{#1}}
\newcommand{\uvett}[1]{\hat{\vett{#1}}}

\begin{document}

\twocolumn[ 

\title{\GH and \IF shifts for Gaussian beams impinging on graphene-coated surfaces}

\author{Simon Grosche, Marco Ornigotti and Alexander Szameit}

\address{
Institute of Applied Physics, Friedrich-Schiller-Universit\"at Jena, \\ Max-Wien-Platz 1, D-07743 Jena, Germany \\
}

\begin{abstract} 
We present a theoretical study of the \GH and \IF shifts for a fundamental Gaussian beam impinging on  a surface coated with a single layer of graphene. We show that the graphene surface conductibility $\sigma(\omega)$ is responsible for the appearance of a giant and negative spatial  \GH shift. 
\end{abstract}

\ocis{240.3695, 260.2110, }

 ] 
When an optical beam impinges upon a surface, nonspecular reflection phenomena may occur, such as the Goos-H\"anchen (GH) \cite{refGH1,refGH2,refGH3} and Imbert-Fedorov (IF) \cite{refIF1,refIF2} shifts, resulting in an effective beam shift at the interface. A comprehensive review on beam shift phenomena can be found in Ref. \cite{res9}. Although Goos and H\"anchen published their work more than 60 years ago \cite{refGH1}, this field of research is still very active, and in the last decades a vast amount of literature has been produced on the subject, resulting not only in a better understanding of the underlying physical principles  \cite{refIF3,refIF4,refIF5,refIF6,refIF7,refIF8,refIF10,res10}, but also in the careful investigation of the effects of various field configurations \cite{beam1,beam2,beam3,beam4,beam5} and reflecting surfaces \cite{surf1,surf2,surf3,surf4} on the GH and IF shifts.

In recent years, on a parallel trail, graphene attracted very rapidly a lot of interest, thanks to its intriguing properties \cite{ref1,ref2}. Its peculiar band structure and the existence of the so-called Dirac cones \cite{ref3}, for example, give the possibility to use graphene as a model to observe QED-like effects such as Klein tunneling \cite{ref5}, Zitterbewegung \cite{ref6}, the anomalous quantum Hall effect \cite{ref8} and the appearance of a minimal conductivity that approaches the quantum limit $e^2/\hbar$ for vanishing charge density \cite{ref9}. In addition, the reflectance and transmittance of graphene are determined by the fine structure constant \cite{ref10}, and a single layer of graphene shows universal absorbance in the spectral range from near-infrared to the visible part of the spectrum \cite{ref11}. 

Among the vast plethora of applications, graphene also proved to be a very interesting system where to observe beam shifts. Very recently, in fact, the occurrence of GH shift in graphene-based structures has been reported, both for light beams (where giant GH shift has been observed \cite{ref13}) and for Dirac fermions \cite{ref14}. Despite all this,  a full theoretical analysis of GH and IF shifts in a graphene-based structure has not been yet carried out.

In this Letter, we therefore present a theoretical analysis of the GH and IF shifts occurring for a monochromatic Gaussian beam impinging onto a glass surface coated with a single layer of graphene. The results of our investigations show on one hand, that the appearance of a giant GH shift is ultimately due to the graphene's surface conductivity $\sigma(\omega)$, and on the other hand, that the presence of the single layer of graphene introduces a dependence of the phases of the reflection coefficients on the incidence angle, thus resulting in a nonzero spatial GH shift also when total internal reflection does not occur.

\begin{figure}[t!]
\centering
\includegraphics[width=0.5\textwidth]{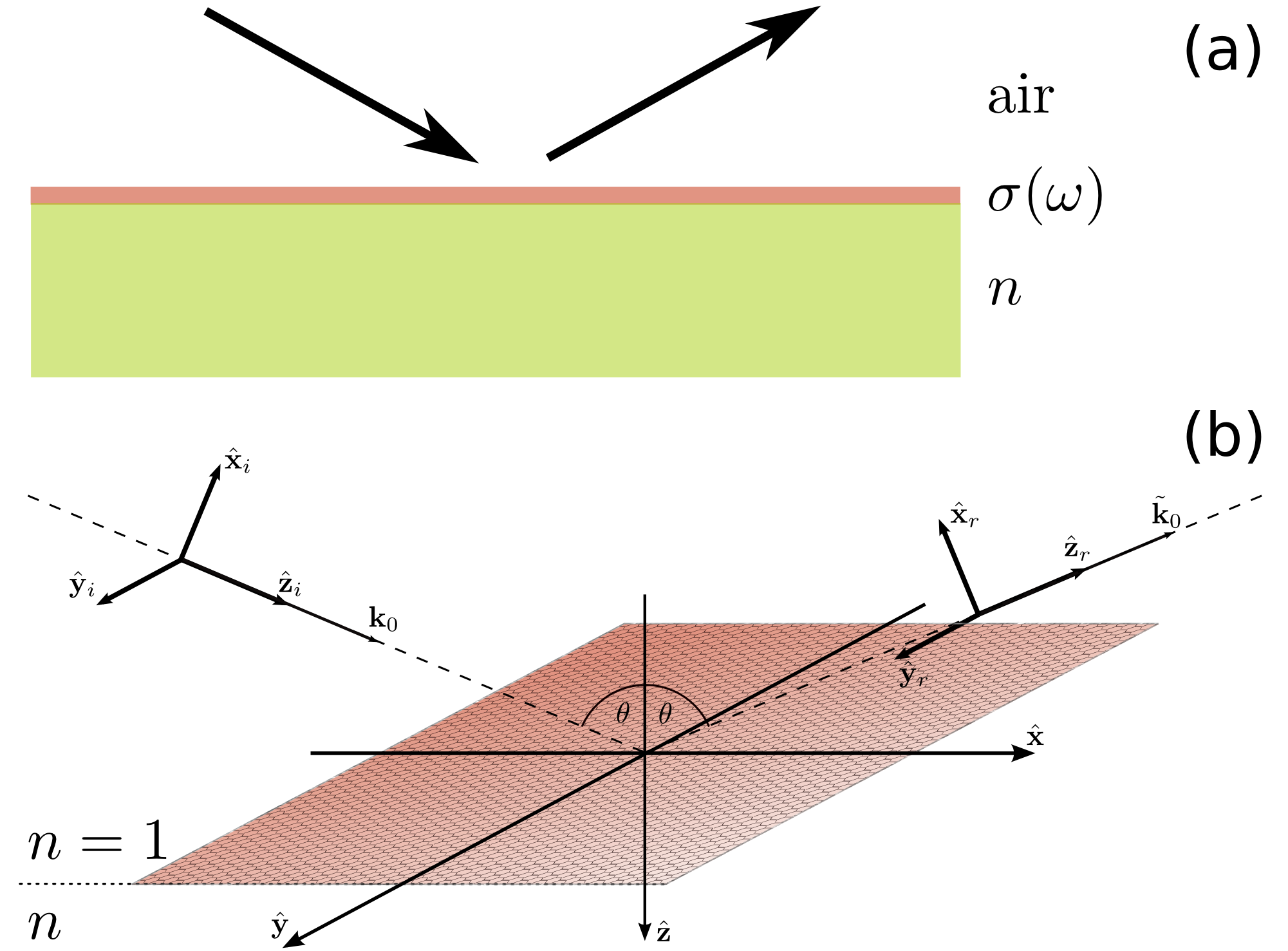}
\caption{(Color online) (a) Schematic representation of the considered surface. The graphene layer (red) is characterized by its surface conductivity $\sigma(\Omega)$, whose explicit expression is given by Eq. \eqref{sigma}. The dielectric substrate (green) is characterized by the refractive index $n$. (b) Geometry of beam reflection at the interface. The single graphene layer is located on the surface at $z=0$. The different Cartesian coordinate systems $K,K_i,K_r$ are shown.}
\label{fig1}
\end{figure}

We start our analysis by considering a monochromatic Gaussian beam with frequency $\omega=ck$ (with $k$ being the vacuum wave number), impinging on a dielectric surface characterized by the refractive index $n$ and coated with a single layer of graphene [Fig. \ref{fig1} (a)]. The graphene layer is characterized by the optical conductivity $\sigma(\omega)$, whose expression can be given in the following dimensionless form \cite{ref3}
\begin{equation}\label{sigma}
\sigma(\Omega)= \mathrm{i}\frac{4 \alpha}{\Omega}+ \pi \alpha \left[\Theta(\Omega-2)+\frac{\mathrm{i}}{2\pi} \ln \frac{(\Omega-2)^2}{(\Omega+2)^2}\right],
\end{equation}
where $\Omega = \hbar \omega/\mu$ is the dimensionless frequency, $\mu$ is the chemical potential, $\alpha\approx1/137$ is the fine structure constant \cite{ref6} and $\Theta(x)$ is the Heaviside step function \cite{nist}. 

According to Fig.\,\ref{fig1}(b), we define three Cartesian reference frames:  the laboratory frame $K=(O,x,y,z)$ attached to the reflecting surface, the (local) incident frame $K_i=(O,x_i,y,z_i)$ attached to the incident beam, and the (local) reflected frame $K_r=(O,x_r,y,z_r)$ attached to the reflected beam. These three reference frames are connected via a rotation of an angle $\theta$ around the $y$ direction \cite{PRA_Aiello}. The reflecting surface is located at $z=0$ , with the $z$-axis pointing towards the interface. With this choice of geometry, the incident beam comes from the region $z<0$ and propagates in the $x$-$z$ plane. 

The electric field in the incident frame can be then written, using its angular spectrum representation \cite{mandelWolf}, as follows:
\begin{equation}
\vett{E}_i(\vett{r})=\sum_{\lambda=1}^2\int d^2K\, \uvett{e}_{\lambda}(U,V,\theta) A_{\lambda}(U,V,\theta)e^{i\vett{k}_i\cdot\vett{r}_i},
\end{equation}
where $d^2K=dUdV$, $\uvett{e}_{\lambda}(U,V,\theta)$ is the local reference frame attached to the incident field \cite{brewster}, $A_{\lambda}(U,V,\theta)=\alpha_{\lambda}(U,V,\theta)A(U,V)$ and $\vett{k}_i\cdot\vett{r}_i=UX_i+VY_i+WZ_i$, 
being $X_i=k_0x_i$ the normalized coordinate in the incident frame. $Y_i$ and $Z_i$ are defined in a similar manner. $\alpha_{\lambda}(U,V,\theta)=\uvett{e}_{\lambda}(U,V,\theta)\cdot\uvett{f}$ accounts for the projection of the beam's polarization $\uvett{f}=f_p\uvett{x}+f_s \uvett{y}$ (normalized according to  $|f_p|^2+|f_s |^2=1$) onto the local basis, and $A(U,V)$ is the beam's spectral amplitude, which here is assumed to be Gaussian, i.e.,
\begin{equation}
A(U,V)=e^{-w_0^2(U^2+V^2)},
\end{equation}
being $w_0^2$ the spot size of the beam. In the remaining of the manuscript, we will consider only well collimated beams, namely the paraxial assumption $U,V\ll 1$ is implicitly understood. 

Upon reflection, the electric field can be then written as follows:
\begin{equation}
\vett{E}_r(\vett{r}_r)=\sum_{\lambda =1}^2 \int d^2K\, \hat{\mathbf{e}}_\lambda (-U,V,\pi-\theta) \tilde{A}_\lambda(U,V,\theta) \mathrm{e}^{\mathrm{i}\vett{k}_r \cdot \vett{r}_r} ,
\label{eq:Er}
\end{equation}
where $\vett{k}_r \cdot \vett{r}_r=-UX_r+VY_r+WZ_r$ and
$\tilde{A}_\lambda(U,V,\theta)= r_\lambda(U,V,\theta) A_\lambda(U,V,\theta)$, with $r_{\lambda}(U,V,\theta)$ being the Fresnel reflection coefficients associated to the single plane wave component of the field \cite{BornWolf}. The minus sign in front of $U$ in  
$\hat{\mathbf{e}}_{\lambda}$, as well as in $\vett{k}_r \cdot \vett{r}_r$ accounts for the specular reflection of the single plane wave component \cite{brewster}.

\begin{figure*}[!t]
\begin{center}
\includegraphics[width=\textwidth]{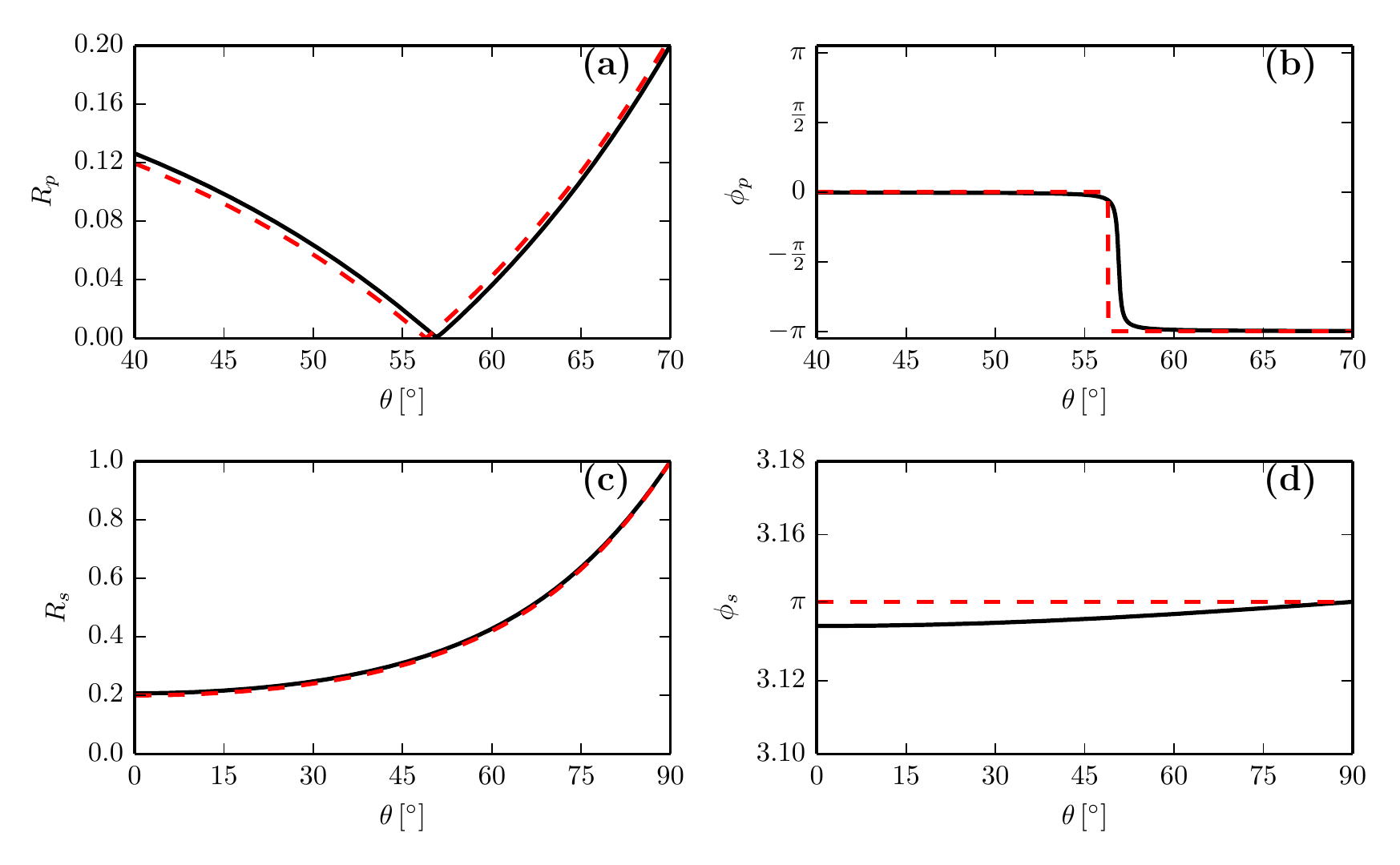}
\caption{Modulus (left column) and phase (right column) of the reflection coefficients $r_{\lambda}=R_{\lambda}\exp{(i\phi_{\lambda})}$ for $p$-polarization (top row) and $s$-polarization (bottom row). In all graphs, the solid black line corresponds to the case of the graphene-coated surface, while the red dashed curve corresponds to the case without graphene coating. The refractive index of the bulk medium is chosen to be $n=1.5$.}
\label{fig2alt}
\end{center}
\end{figure*}

The presence of a single layer of graphene deposited on the dielectric surface modifies its reflection coefficients as follows \cite{transferMatrix}:
\begin{subequations}
\label{reflection}
\begin{align}
r_s(\theta)&=\frac{\cos\theta-\sqrt{n^2-\sin^2\theta}-\sigma(\Omega)}{\cos\theta+\sqrt{n^2-\sin^2\theta}+\sigma(\Omega)}
,\\
r_p(\theta)&=\frac{n^2\cos\theta-\sqrt{n^2-\sin^2\theta}\left[1-\sigma(\Omega)\cos\theta\right]}{n^2\cos\theta+\sqrt{n^2-\sin^2\theta}\left[1+\sigma(\Omega)\cos\theta\right] } ,
\end{align}
\end{subequations}
where $\theta$ is the incident angle, $n$ is the refractive index of the dielectric medium and $\sigma(\Omega)$ is the graphene's surface conductivity, as defined by Eq. \eqref{sigma}.  The modulus $R_{\lambda}$  and phase 
$\phi_{\lambda}$ of the reflection coefficients $r_{\lambda}=R_{\lambda}e^{i\phi_{\lambda}}$ (with $\lambda\in\{p,s\}$) are shown in Fig. \ref{fig2alt}, together with the correspondent quantities for the case of a simple dielectric surface without the graphene coating. While the presence of the graphene layer does not modify significatively  $R_{\lambda}$ for neither $p$- or $s$-polarization (as it appears clear from Figs. \ref{fig2alt}(a) and (c), respectively), the change induced in the phases $\phi_{\lambda}$ of the reflection coefficients is considerable. For a normal air-glass interface, in fact, we have $\partial\phi_{\lambda}/\partial\theta=0$ being $\theta$ the angle of incidence. Here, instead, we have $\partial\phi_{\lambda}/\partial\theta\neq 0$. 
A closer inspection of Eqs. \eqref{reflection}, moreover, reveals that such a novel $\theta$-dependence of the phases  $\phi_{\lambda}$ is entirely due to the graphene conductivity $\sigma(\Omega)$.

To compute the GH and IF shifts, we calculate the center of mass of the intensity distribution in the reflected frame, namely \cite{brewster}
\begin{equation}
\langle \vett{R}\rangle=\frac{\iint\limits_{-\infty}^{+\infty} \vett{R} |\mathbf{E}_r|^2 \mathrm{d}X_r \mathrm{d}Y_r}{\iint\limits_{-\infty}^{+\infty} |\mathbf{E}_r|^2 \,\mathrm{d}X_r \mathrm{d}Y_r}
= \langle X_r\rangle\uvett{X}_r+\langle Y_r\rangle\uvett{y}_r,
\label{eq:GHIFshiftDef}
\end{equation}
where $\vett{R}=(X_r,Y_r)^T$. Spatial ($\Delta$) and angular ($\Theta$) GH and IF shifts are then defined as follows:
\begin{subequations}
\begin{align}
\Delta_{GH}&=\langle  X_r\rangle \Big|_{z=0},& \quad
\Theta_{GH}&=\frac{\partial\langle  X_r\rangle}{\partial z},\\
\Delta_{IF}&=\langle  Y_r\rangle \Big|_{z=0},& \quad
\Theta_{IF}&=\frac{\partial\langle  Y_r\rangle}{\partial z}.
\end{align}
\end{subequations}
The explicit expressions of the GH and IF shifts for a fundamental Gaussian beam read, according to \cite{poly1}, as follows:
\begin{subequations}\label{shifts}
\begin{align}
\Delta_{GH} &=w_p\frac{\partial\phi_p}{\partial\theta}+w_s\frac{\partial\phi_s}{\partial\theta},\\
\Delta_{IF} &=-\cot\theta\Big[\frac{w_pa_s^2+w_sa_p^2}{a_pa_s}\sin\eta\nonumber\\
&+2\sqrt{w_pw_s}\sin(\eta-\phi_p+\phi_s)\Big],\\
\Theta_{GH}&=-\left(w_p\frac{\partial\ln R_p}{\partial\theta}+w_s\frac{\partial\ln R_s}{\partial\theta}\right),\\
\Theta_{IF}&=\frac{w_pa_s^2-w_sa_p^2}{a_pa_s}\cos\eta\cot\theta,
\end{align}
\end{subequations}
where $f_p=a_p$, $f_s=a_s\exp{(i\eta)}$ and $w_{\lambda}=a_{\lambda}^2R_{\lambda}^2/(a_p^2R_p^2+a_s^2R_s^2)$ (where $\lambda\in\{p,s\}$) is the fractional energy contained in each polarization state. 

As suggested by Figs. \ref{fig2alt}(a) and (c),  the changes in $R_{\lambda}$ introduced by the graphene layer are negligible. We therefore expect to observe no changes in the angular shifts $\Theta_{GH}$ and $\Theta_{IF}$, as they are functions of  $R_{\lambda}$ solely. The spatial shifts $\Delta_{GH}$ and $\Delta_{IF}$, on the other hand, contain a dependence on the phases $\phi_{\lambda}$, and they are therefore affected by the presence of the graphene coating. Let us first discuss the IF shift. In this case $\phi_p-\phi_s$ is very close to $\pi$ [Figs. \ref{fig2alt}(b) and (d)],  and the resulting spatial shift $\Delta_{IF}$ will be \emph{nonzero} (but very small) even for linear polarization, in contrast to the case without graphene.
 
 More interesting is the case of the spatial GH shift. For a normal air-dielectric interface, one has  $\partial\phi_{\lambda}/\partial\theta=0$ and therefore, according to Eq. (8a), $\Delta_{GH}=0$. It is in fact well known since the pioneering work of Goos and H\"anchen \cite{refGH1}, that $\Delta_{GH}\neq 0$ occurs only in total internal reflection, where $R_{\lambda}=1$ and $\partial\phi_{\lambda}/\partial\theta\neq0$. For the case of a graphene-coated surface, on the other hand, the phase $\phi_{\lambda}$ varies with $\theta$ for both $s$- and $p$-polarizations, as Figs. \ref{fig2alt} (b) and (d), respectively, show. In this case, then, we observe a \emph{nonzero} spatial GH shift even without total internal reflection.
 
\begin{figure*}[t!]
\begin{center}
\includegraphics[width=1.\textwidth]{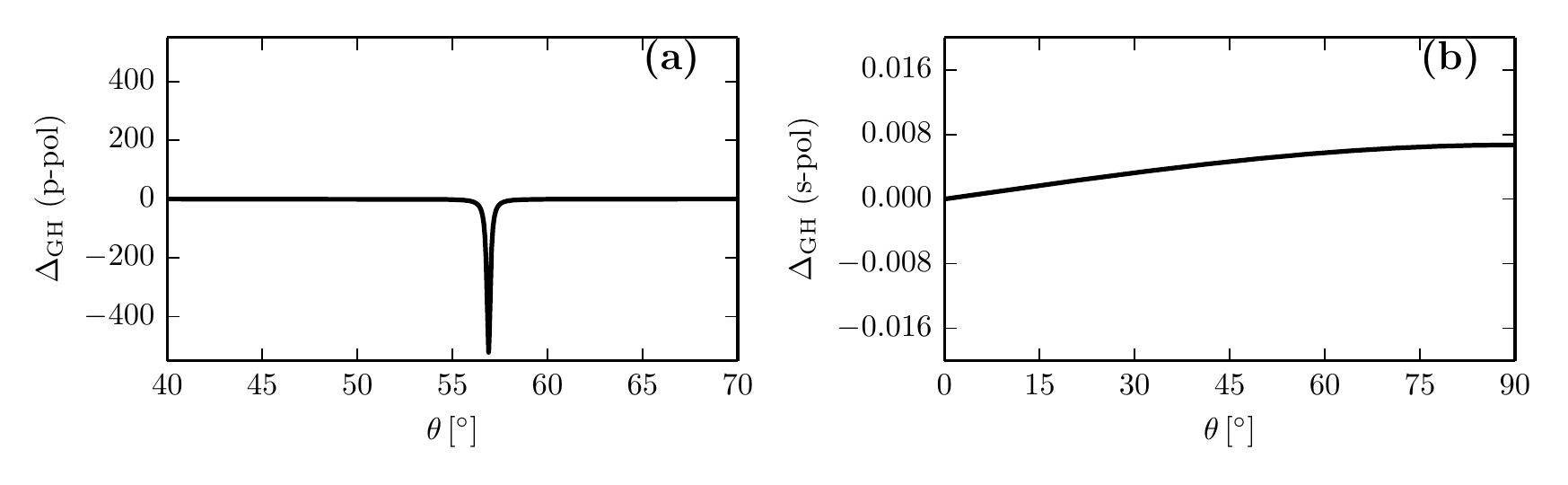}
\caption{Spatial GH shift $\Delta_{GH}$ for (a) $p$- polarization and (b) $s$-polarization for a graphene-coated surface. Since $\partial\phi_{\lambda}/\partial\theta\neq 0$, in both cases $\Delta_{GH}\neq 0$. In particular, since $\phi_p$ varies very rapidly with $\theta$ in the vicinity of the Brewster angle, the corresponding spatial shift for $p$-polarization [Panel (a)] is giant in modulus, and negative due to the fact that $\phi_p$ varies from $0$ to $-\pi$ [See Fig. \ref{fig2alt}(b)].}
\label{fig3}
\end{center}
\end{figure*}

 The spatial GH shift $\Delta_{GH}$ occurring at a graphene-coated dielectric surface is depicted in Fig. \ref{fig3}(a) and (b) for $p$- and $s$-polarization, respectively. As can be seen, for both polarizations we have $\Delta_{GH}\neq 0$ although no total internal reflection takes place. In particular, $\phi_p$ varies very rapidly from $0$ to $-\pi$ in the vicinity of the Brewster angle $\theta_B$. This corresponds to a giant and negative spatial GH shift. On the other hand, $\phi_s$ varies very smoothly with $\theta$, thus resulting in a \emph{nonzero} (but very small) spatial GH shift for $s$-polarization.
 
In conclusion, we have presented a detailed theoretical analysis of GH and IF shifts of a Gaussian beam impinging onto a graphene-coated dielectric surface. Our analysis revealed that the main effect of the graphene layer is to introduce, through its surface conducibility $\sigma(\omega)$, a dependence of the phases $\phi_{\lambda}$ of the reflection coefficients  on the incident angle $\theta$. This, ultimately, reflects in the appearance of a \emph{nonzero} spatial GH and IF shifts. In particular a giant and negative spatial GH shift in the vicinity of the Brewster's angle for $p$-polarization has been predicted, in ageeement with the recently published experimental results \cite{ref13}.
The authors thank the German Ministry of Education and Science (ZIK 03Z1HN31) for financial support.

\clearpage

\begin{thebibliography}{99}

\bibitem{refGH1}F. Goos and H. H$\mathrm{\ddot{a}}$nchen, Ann. Phys. \textbf{1}, 333 (1947).
\bibitem{refGH2}K. Artmann, Ann. Phys. \textbf{2}, 87 (1948).
\bibitem{refGH3}K. W. Chiu and J. J. Quinn, Am. J. Phys. \textbf{40}, 1847 (1972)
\bibitem{refIF1}F.  I. Fedorov, Dokl. Akad. Nauk SSSR \textbf{105}, 465 (1955).
\bibitem{refIF2}C. Imbert, Phys. Rev. D \textbf{5}, 787 (1972).
\bibitem{res9}K. Y. Bliokh and A. Aiello, J. Opt. \textbf{15}, 014001 (2013).

\bibitem{refIF3}F. Pillon, H. Gilles and S. Girard, Appl. Opt. \textbf{43}, 1863 (2004).
\bibitem{refIF4}H. Schilling, Ann. Phys. (Berlin) \textbf{16}, 122 (1965).
\bibitem{refIF5}M. A. Player,  J. Phys. A: Math. Gen. \textbf{20}, 3667 (1987).
\bibitem{refIF6}V. G. Fedoseyev,  J. Phys. A: Math. Gen \textbf{21}, 2045 (1988).
\bibitem{refIF7}V. S. Liberman and B. Y. Zel'dovich, Phys. Rev. A \textbf{46}, 5199 (1992).
\bibitem{refIF8}K. Y. Bliokh and Y. P. Bliokh, Phys. Rev. Lett. \textbf{96}, 073903 (2006).
\bibitem{refIF10}A. Aiello and J. P. Woerdman, Opt. Lett. \textbf{33}, 1437 (2008).
\bibitem{res10}O. Hosten and P. Kwiat, Science \textbf{319}, 787 (2008).

\bibitem{beam1}P.T. Leung, C. W. Chen and H. -P. Chiang, Opt. Commun. \textbf{276}, 206 (2007).
\bibitem{beam}M.  Merano, A. Aiello, G. W. 't Hooft, M. P. von Exter, E. R. Eliel and J. P. Woerdman, Opt. Expr. \textbf{15}, 15928 (2007).
\bibitem{beam3}T. Tamir, J. Opt. Soc. Am. A \textbf{3}, 558 (1986).
\bibitem{beam4}G. D. Landry and T. A. Maldonado, Appl. Opt. \textbf{35}, 5870 (1996).
\bibitem{beam5}M.Ornigotti, A. Aiello and C. Conti, Opt. Lett. \textbf{40}, 558 (2015).

\bibitem{surf1}S. Kozaki and H. Sakurai, J. Opt. Soc. Am. \textbf{68}, 508 (1978).
\bibitem{surf2}D. Golla and S. Dutta Gupta, arXiv:1011.3968v1.
\bibitem{surf3}M. Merano, N. Hermosa, J.P. Woerdman and A. Aiello, Phys. Rev. A \textbf{82}, 023817 (2010).
\bibitem{surf4}A. Aiello and J. P. Woerdman, Opt. Lett. \textbf{36}, 543 (2010).


\bibitem{ref1} K. S. Novoselov, A. K. Geim, S. V. Morozov, D. Jiang, M. I. Katsnelson, I. V. Grigorieva, S. V. Dubonos and A. A. Firsov, Nature \textbf{438}, 197 (2005).
\bibitem{ref2}A. H. Castro Neto, F. Guinea, N. M. R. Peres, K. S. Novoselov, and A. K. Geim, Rev. Mod. Phys. \textbf{81}, 109 (2009).
\bibitem{ref3}M. I. Katsnelson, \emph{Graphene: Carbon in Two Dimensions} (Cambridge University Press, 2012).
\bibitem{ref5}A. Calogeracos, N. Dombey, Contemp. Phys. \textbf{40}, 313 (1999).
\bibitem{ref6}C. Itzykson and J. B. Zuber, \emph{Quantum Field Theory} (Dover, 2006).
\bibitem{ref8}Y. Zhang, Y. W. Tan, H. L. Stormer and P. Kim, Nature \textbf{438}, 201 (2005).
\bibitem{ref9}M. I. Katsnelson, Eur. Phys. J. B. \textbf{51}, 157 (2006).
\bibitem{ref10}R. R. Nair, P. Blake, A. N. Grigorenko, K. S. Novoselov, T. J. Booth, T. Stauber, N. M. R. Peres and A. K. Geim,  Science \textbf{320}, 1308 (2008).
\bibitem{ref11}T. Stauber, N. M. R. Peres, and A. K. Geim,  Phys. Rev. B \textbf{78}, 085432 (2008).



\bibitem{ref13}X. Li, P. Wang, F. Xing, X. D. Chen, Z. B. Liu, and J. G. Tian, Opt. Lett. \textbf{39}, 5574 (2014).
\bibitem{ref14}A. Jellala, I. Redouanic, Y. Zahidic and H. Bahloulia, Physica E \textbf{58}, 30 (2014).
\bibitem{nist}\emph{Digital Library of Mathematical Functions}, http://dlmf.nist.gov, National Institute of Standard and Technology (2010).
\bibitem{mandelWolf} L. Mandel and E. Wolf, \emph{Optical Coherence and Quantum Optics}, (Cambridge University Press, New York, 1995).

\bibitem{PRA_Aiello} M. Merano A. Aiello and J. P. Woerdman, Phys. Rev. A \textbf{80}, 061801(R) (2009).
\bibitem{brewster}A. Aiello and J. P. Woerdman,  arXiv:0903.3730v2 [physics.optics].
\bibitem{BornWolf}M. Born and E. Wolf, \emph{Principles of Optics}, 7th edition (Cambridge University Press, 2003). 
\bibitem{transferMatrix}T. Zhan, X. Shi, Y. Dai, X. Liu and J. Zi,  J. Phys.: Conden. Matt. \textbf{25},  215301 (2013).
\bibitem{poly1} M. Ornigotti and A. Aiello, J. Opt. \textbf{15}, 014004 (2013).
\end{thebibliography}
\end{document}